\documentclass[11pt]{article}
\usepackage{amsmath}
\usepackage{amssymb}
\usepackage{amsthm}
\usepackage{xcolor}
\usepackage{caption}
\usepackage{calc}
\usepackage{hyperref}
\usepackage{graphicx}
\usepackage{wrapfig}
\usepackage{enumitem}
\usepackage{cite}
\usepackage{geometry}
\usepackage{algorithm}
\usepackage[noend]{algpseudocode}
\newcounter{algsubstate}
\renewcommand{\thealgsubstate}{\alph{algsubstate}}

\begin{document}

\title{Network analysis using Krylov subspace trajectories}

\author{H. Robert Frost$^{1}$}
\date{}
\maketitle
\begin{center}
\textit{
$^1$Department of Biomedical Data Science\\
Geisel School of Medicine \\
Dartmouth College \\
Hanover, NH 03755, USA \\
rob.frost@dartmouth.edu
}
\end{center}

\begin{abstract}
We describe a set of network analysis methods based on the rows of the Krylov subspace matrix computed from a network adjacency matrix via power iteration using a non-random initial vector. We refer to these node-specific row vectors as Krylov subspace trajectories. While power iteration using a random initial starting vector is commonly applied to the network adjacency matrix to compute eigenvector centrality values, this application only uses the final vector generated after numerical convergence. Importantly, use of a random initial vector means that the intermediate results of power iteration are also random and lack a clear interpretation. To the best of our knowledge, use of intermediate power iteration results for network analysis has been limited to techniques that leverage just a single pre-convergence solution, e.g., Power Iteration Clustering. In this paper, we explore methods that apply power iteration with a non-random inital vector to the network adjacency matrix to generate Krylov subspace trajectories for each node. These non-random trajectories provide important information regarding network structure, node importance, and response to perturbations. We have created this short preprint in part to generate feedback from others in the network analysis community who might be aware of similar existing work.
\end{abstract}

\section{Krylov subspace}\label{sec:krylov}

The order $m$ Krylov subspace \cite{Krylov, 10007764} for a square diagonalizable $p \times p$ matrix $\mathbf{X}$ and length $p$ vector $\mathbf{v}$ corresponds to the sequence of $m$ vectors generated by the product of $\mathbf{v}$ with powers of $\mathbf{X}$ ranging from 0 to $m-1$, i.e. $\{\mathbf{v}, \mathbf{Xv}, \mathbf{X^2v}, ... , \mathbf{X^{m-1}v}\}$. This sequence of vectors can be used to form the columns of a $p \times m$ Krylov subspace matrix $\mathbf{K}$:

\begin{equation}\label{eqn:krylov_matrix}
\mathbf{K} = \begin{bmatrix}
\mathbf{v} & \mathbf{Xv} & \mathbf{X^2v} & \cdots & \mathbf{X^{m-1}v} \\
\end{bmatrix}
\end{equation}

\noindent We refer to the row vectors of $\mathbf{K}$ as Krylov subspace trajectories with the elements of the trajectory for row $i$ defined by the recursive structure:

\begin{equation}\label{eqn:krylov_trajectory}
\mathbf{t_i} = \{ v_i, \sum_{j=1}^p x_{i,j} v_j, \sum_{j=1}^p x_{i,j} (\sum_{k=1}^p x_{i,k} v_k),  ...\}
\end{equation}

\noindent For practical numerical applications, the normalized Krylov subspace matrix is used where each column vector is normalized to unit length:

\begin{equation}\label{eqn:krylov_matrix}
\mathbf{K}_n = \begin{bmatrix}
\frac{\mathbf{v}}{||\mathbf{v}||_2} & \frac{\mathbf{Xv}}{||\mathbf{Xv}||_2} & \frac{\mathbf{X^2v}}{||\mathbf{X^2v}||_2} & \cdots & 
\frac{\mathbf{X^{m-1}v}}{||\mathbf{X^{m-1}v}||_2} \\
\end{bmatrix}
\end{equation}

\noindent In the remainder of this paper, references to the Krylov subspace matrix and associated trajectories will assume that the normalized version is being used. 

The Krylov subspace has a number of useful properties and is leveraged in a wide range of numerical linear algebra techniques. In particular, the columns of $\mathbf{K}$ will generally converge to the eigenvector of $\mathbf{X}$ associated with the largest eigenvalue as long as $\mathbf{v}$ is not orthogonal to this principal eigenvector. The columns of $\mathbf{K}$ are also linearly independent so long as $m \leq r + 1$ where $r$ is the rank of $\mathbf{X}$. These properties support techniques like Arnoldi iteration \cite{arnoldi:hal-01712943}, which computes the eigenvectors and eigenvalues of $\mathbf{X}$ using the Krylov subspace with incremental orthogonalization. Other Krylov subspace methods include Lanczos iteration \cite{Lanczos:1950zz}, the conjugate gradient method \cite{Hestenes&Stiefel:1952} and the minimal residual method \cite{doi:10.1137/0712047}.

\subsection{Krylov subspace and power iteration}\label{sec:krylov_power_iteration}

\noindent The Krylov subspace has a direct correspondence with the incremental results from the method of power iteration \cite{MisesPraktischeVD}, whose common realization is defined by Algorithm \ref{alg:power_iteration}. In particular, the sequence of vectors $\{\mathbf{v}_0, \mathbf{v}_1, ..., \mathbf{v}_i\}$ generated by power iteration correspond to the order $i+1$ normalized Krylov subspace for $\mathbf{X}$ and $\mathbf{v}_0$.
Note that the specification of power iteration in Algorithm \ref{alg:power_iteration} does not include features important for many numerical applications, e.g, checks to ensure that $\mathbf{X}_i$ is well conditioned, stochastic initialization of $\mathbf{v}_0$, use of methods like accelerated stochastic power iteration \cite{pmlr-v84-xu18a} to improve computational performance, alternate stopping conditions, etc..

\begin{algorithm}
\caption{Power iteration}
\label{alg:power_iteration}
\hspace*{\algorithmicindent} \textbf{Input:} 
\begin{itemize}
\setlength\itemsep{0em}
\item A $p \times p$ matrix $\mathbf{X}$.
\item A length $p$ vector $\mathbf{v}_0$ used to initialize the iteration.
\item Positive integer $maxIter$ that represents the maximum number of iterations.
\item Positive real number $tol$ that represents the stopping criteria as the proportional change in the estimated principal eigenvalue between iterations.
\end{itemize}
\hspace*{\algorithmicindent} \textbf{Output:} 
\begin{itemize}
\setlength\itemsep{0em}
\item Estimated eigenvector $\mathbf{v}$ of $\mathbf{X}$ associated with the largest eigenvalue $\lambda$.
\item Estimated principal eigenvalue $\lambda$.
\item Number of iterations completed.
\end{itemize}
\begin{algorithmic}[1]
\For{$i \in \{1,...,maxIter\}$}
\State $\mathbf{v}_{i} = \mathbf{X} \mathbf{v}_{i-1}$
 \Comment{Update $\mathbf{v}_i$}

\State $\mathbf{v}_i = \mathbf{v}_i/||\mathbf{v}_i||_2$
 \Comment{Normalize $\mathbf{v}_i$ to unit length}
 
\State $\lambda_{i} = \mathbf{v}_{i}^T \mathbf{X} \mathbf{v}_i$
 \Comment{Compute principal eigenvalue estimate}
  
\If{$i > 1$}
 \State $\delta = (| \lambda_{i-1} - \lambda_{i}| )/\lambda_{i}$ 
   \Comment{Compute proportional change in eigenvalue estimate}
  \If{$\delta < tol$}
    \State \textbf{break} \Comment{If proportion change is less than $tol$, exit}
  \EndIf
\EndIf
\EndFor
\Return {$\mathbf{v}_{i}, \lambda_{i}, i$}

\end{algorithmic}
\end{algorithm}

\clearpage

\subsection{Existing uses of the Krylov subspace for network analysis}\label{sec:krylov_network}

In the network analysis space, the Krylov subspace is most commonly generated through the application of power iteration to the network adjacency matrix, $\mathbf{A}$, which, for an edge-weighted network defined over $p$ nodes is the $p \times p$ matrix:

\begin{equation} \label{eqn:adj_matrix}
\mathbf{A} = \begin{bmatrix}
a_{1,1} & \cdots & a_{1,p} \\
\vdots & \ddots & \vdots \\
a_{p,1} & \cdots & a_{p,p}
\end{bmatrix}
\end{equation}

\noindent where $a_{i,j}$ holds the weight of the edge between nodes $i$ and $j$ or 0 if no such edge exists. Self-edges are represented by the diagonal elements. If the network is undirected, $a_{i,j} = a_{j,i}$ and $\mathbf{A}$ is symmetric; if the network is directed, then $a_{i,j}$ and $a_{j,i}$ capture distinct edges and $\mathbf{A}$ is asymmetric. Power iteration is normally applied to the network adjacency matrix using a random $\mathbf{v}_0$ to compute eigenvector centrality values (or values for a related centrality measure)  \cite{Newman:2010ve}. Eigenvector centrality models node importance as the weighted sum of the importance of adjacent nodes and therefore leads to the  following eigenvalue problem:
 
\begin{equation} \label{eqn:standard_ec}
\mathbf{A} \mathbf{v} = \lambda \mathbf{v}
\end{equation}
 
\noindent Specifically, the eigenvector centrality for node $n$ is equal to element $n$ of the eigenvector $\mathbf{v}$ corresponding to the largest eigenvalue. When the network is strongly connected, the matrix $\mathbf{A}$ is irreducible and the Perron-Frobenius theorem \cite{Perron:1907ua} guarantees that there is a unique largest real eigenvalue with a corresponding eigenvector that has strictly positive elements. 

While this application of power iteration generates the $k+1$ order Krylov subspace of $\mathbf{A}$, where $k$ is the number of iterations until numerical convergence, computation of eigenvector centrality only uses the final vector. In addition, the fact that a random $\mathbf{v}_0$ is almost always employed means the generated trajectories are also random and therefore lack a clear interpretation. Other network analysis techniques based on the Krylov subspace of $\mathbf{A}$ (or a related matrix) also just use the output from a single iteration of the power method. These techniques include the propagation of node information in Graph Convolution networks via a single power method iteration \cite{ye2023graph} and the Power Iteration Clustering method \cite{10.5555/3104322.3104406} that performs a truncated power iteration using a random initial vector on a normalized affinity matrix and then uses the last eigenvector estimate to perform node clustering via the k-means algorithm.

\section{Krylov subspace trajectories of the network adjacency matrix}

In this paper, we are interested in the Krylov subspace associated with a network adjacency matrix $\mathbf{A}$ and non-random initial vector $\mathbf{v}_0$. Generation of this subspace using the fixed initial vector $\mathbf{v}_0 = \{1/\sqrt{p},...,1/\sqrt{p}\}$ is specified by Algorithm \ref{alg:krylov_gen}. 

\begin{algorithm}
\caption{Krylov subspace of network adjacency matrix}
\label{alg:krylov_gen}
\hspace*{\algorithmicindent} \textbf{Input:} 
\begin{itemize}
\setlength\itemsep{0em}
\item A $p \times p$ adjacency matrix $\mathbf{A}$ for a potentially directed, edge-weighted network with $p$ nodes.
\item Positive integer $maxIter$ that represents the maximum number of power iterations.
\item Positive real number $tol$ that represents the stopping criteria as the proportional change in the estimated principal eigenvalue between iterations.
\end{itemize}
\hspace*{\algorithmicindent} \textbf{Output:} 
\begin{itemize}
\setlength\itemsep{0em}
\item The order $i+1$ Krylov subspace matrix $\mathbf{K}$ where $i$ is the number of completed power iterations.
\end{itemize}
\textbf{Notation:}
\begin{itemize}[noitemsep,topsep=0pt]
\item Let $\mathbf{X}[]$ represent a subsetting of the matrix $\mathbf{X}$ with $\mathbf{X}[i,j]$ the element in the $i^{th}$ row and $j^{th}$ column, $\mathbf{X}[i,]$ the $i^{th}$ row, $\mathbf{X}[,j]$ the $j^{th}$ column, and $\mathbf{X}[\mathbf{r}, \mathbf{c}]$ the submatrix containing rows with indices in $\mathbf{r}$ and columns with indices in $\mathbf{c}$.
\end{itemize}
\begin{algorithmic}[1]
\State $\mathbf{K} \in \mathbb{R}^{p \times maxIter+1}$
	\Comment{Initialize $\mathbf{K}$ as a $p \times maxIter+1$ matrix}
\State $\mathbf{K}[,1] = \{1/\sqrt{p}, ... , 1/\sqrt{p}\}$
	\Comment{Set first column of $\mathbf{K}$ to unit length vector with all values $1/\sqrt{p}$}
\For{$i \in \{1,...,maxIter\}$}
\State $\mathbf{K}[,i+1] = \mathbf{X} \mathbf{K}[,i]$
 \Comment{Compute $i+1$ column of $\mathbf{K}$}
\State $\mathbf{K}[,i+1] = \mathbf{K}[,i+1]/||\mathbf{K}[,i+1]||_2$
 \Comment{Normalize column to unit length}
\State $\lambda_{i} = \mathbf{K}[,i+1]^T \mathbf{X} \mathbf{K}[,i+1]$
 \Comment{Compute principal eigenvalue estimate}
  
\If{$i > 1$}
 \State $\delta = (| \lambda_{i-1} - \lambda_{i}| )/\lambda_{i}$ 
   \Comment{Compute proportional change in eigenvalue estimate}
  \If{$\delta < tol$}
    \State \textbf{break} \Comment{If proportion change is less than $tol$, exit}
  \EndIf
\EndIf
\EndFor
\Return {$\mathbf{K}[,\{1,...,i+1\}]$}

\end{algorithmic}
\end{algorithm}

\begin{figure}[t]
\begin{center}
\includegraphics[width=0.75\textwidth]{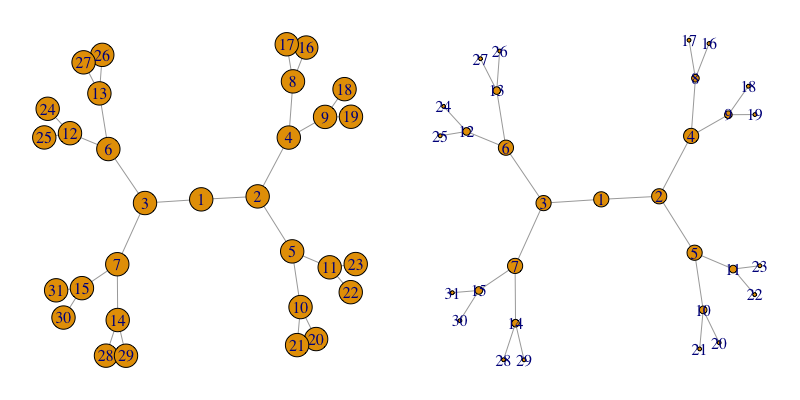}
\end{center}
\caption{Example tree network. The left panel visualizes the network with equal sized nodes. The right panel visualizes with node size based on eigenvector centrality.}
\label{fig:tree}
\end{figure}
\begin{figure}[h]
\begin{center}
\includegraphics[width=0.75\textwidth]{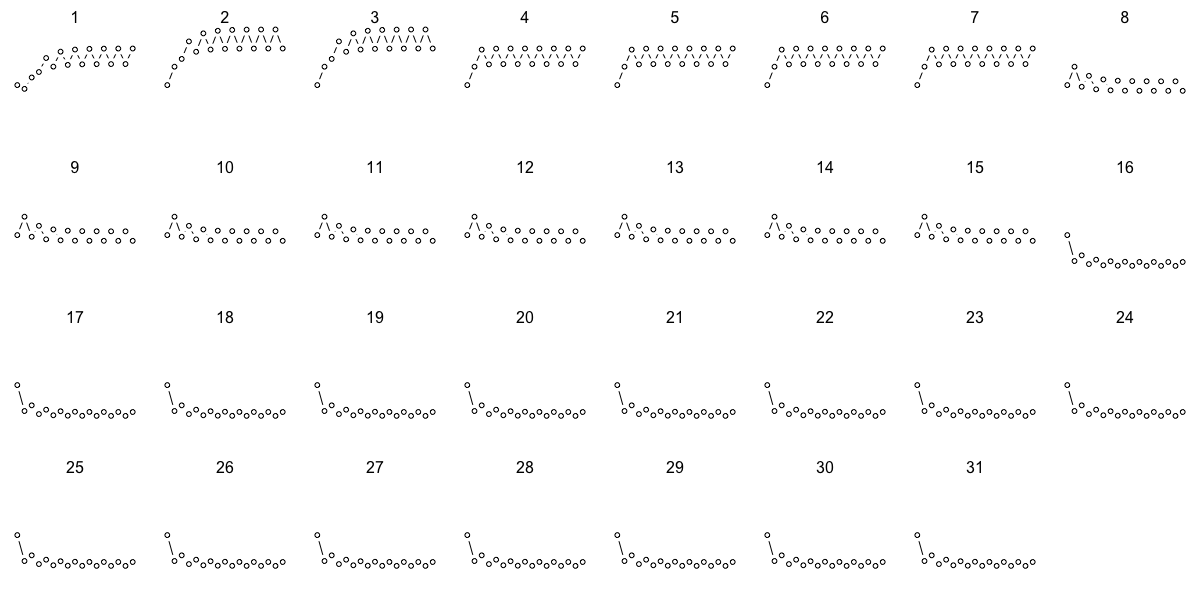}
\end{center}
\caption{Node-specific Krylov trajectories for the tree network visualized in Figure \ref{fig:tree}.}
\label{fig:tree_krylov}
\end{figure}

To illustrate an example of a non-random network Krylov subspace, Algorithm \ref{alg:krylov_gen} was applied to an unweighted and undirected tree network (visualized in Figure \ref{fig:tree}) that was generated with a total of 31 nodes and two child nodes per non-leaf node. For this network, Algorithm \ref{alg:krylov_gen} converged to an estimated eigenvalue of 2.4 after 16 iterations. The order 17 Krylov subspace trajectories of the 31 nodes in this network are visualized in Figure \ref{fig:tree_krylov}. An interesting feature of the trajectories for this tree graph is the presence of oscillations. For this particular network, the power iteration method (as defined in Algorithms 1 and 2) with a uniform starting vector does not in fact converge to the largest eigenvector/eigenvalue pair but instead halts at an eigenvalue estimate of 2.4 (the true largest eigenvalue is 2.4495) with alternating eigenvector estimates $\mathbf{v}_a$ and $\mathbf{v}_b$, i.e., $\mathbf{A}\mathbf{v}_a = \alpha \mathbf{v}_b$, $\mathbf{A}\mathbf{v}_b = \beta \mathbf{v}_a$, and $\mathbf{v}_a^T \mathbf{A} \mathbf{v}_a = \mathbf{v}_b^T \mathbf{A} \mathbf{v}_b = 2.4$). For the more complex edge-weighted networks associated with experimental data, power iteration is unlikely to encounter such erroneous convergence behavior. Although this behavior makes a simple implementation of power iteration ineffective for eigenvector centrality computation on this network, the generated Krylov subspace trajectories still contain important information about network structure. The utility of these trajectories is explored in more detail in Sections \ref{sec:clustering}-\ref{sec:importance}.

\subsection{Krylov subspace matrix and network paths}\label{sec:paths}

The Krylov subspace matrix $\mathbf{K}$ associated with a network adjacency matrix $\mathbf{A}$ has an interesting intepretation in terms of network paths. In particular, the $r^{th}$ column of $\mathbf{K}$ takes the form:

\begin{equation} \label{eqn:krylov_column}
\mathbf{K}[,r] = \mathbf{A}^r \mathbf{v}_0
\end{equation}

\noindent The association with network paths stems from the fact that the elements of $\mathbf{A}^r$ represent the weighted count of paths of length $r$ between each node pair, i.e., element $i,j$ of $\mathbf{A}^r$ holds the number of paths of length $r$ between node $i$ and node $j$ adjusted by the product of the edge weights. For example, if there is only a single path of length 2 between nodes $i$ and $j$ and the edge weights along this path are $\alpha$ and $\beta$, then element $i,j$ of $\mathbf{A}^2$ equals $\alpha \beta$. This implies that element $i$ of the column $\mathbf{K}[,r]$ is proportional to the weighted count of all paths of length $r$ between node $i$ and all other nodes in the network. For the normalized Krylov subspace matrix $\mathbf{K}_n$ defined in Equation \ref{eqn:krylov_matrix}, the elements reflect the weighted count of paths of length $r$ starting from each node relative to the weighted counts for all other nodes. Krylov subspace trajectories can therefore capture changes in the relative weighted path count for a given node as the path length increases. If the trajectory for a node has a monotonically increasing or decreasing trend, this indicates that the weighted path count for the node exhibits a steady increase or decrease relative to other nodes. By contrast, if the trajectory for a node contains oscillations, this indicates that the weighted path count for that node fluctuates in value relative to other nodes as path length increases.

\subsection{Krylov subspace matrix and Leicht, Holme and Newman node similarity}\label{sec:lhn_similarity}

Krylov subspace trajectories also have a close association with the version of node similarity proposed by Leicht, Holme and Newman \cite{PhysRevE.73.026120}. The Leicht, Holme and Newman (LHN) measure of node similarity is based on the recursive model that a node $i$ is similar to a node $j$ if $i$ has a neighbor node $k$ that is itself similar to $j$. If these similarity values are held in a $p \times p$ matrix $\mathbf{S}$, then a simple representation of this model defines the similarity between nodes $i$ and $j$ as:

\begin{equation}\label{eqn:lhn_simple}
\mathbf{S}[i,j] = \phi \sum_{k} \mathbf{A}[i,k] \mathbf{S}[k,j] + \psi \delta_{i,j}
\end{equation}

\noindent where $\delta_{i,j}$ is Kronecker's function and the network is assumed to be undirected, unweighted and without self-loops. According to this equation, a node has a similarity of $\psi$ with itself and the similarity between distinct nodes is a function of the overall network topology. Equation \eqref{eqn:lhn_simple} can be represented in matrix format as:

\begin{equation}\label{eqn:lhn_matrix}
\mathbf{S} = \phi \mathbf{A} \mathbf{S} + \psi \mathbf{I}
\end{equation}

\noindent Adjustments to ensure series convergence and to account for the expected number of paths given node degrees leads to the official LHN similarity equation:

\begin{equation}\label{eqn:lhn}
\mathbf{DSD} = \frac{\alpha}{\lambda_1} \mathbf{ADSD} + \mathbf{I}
\end{equation}

\noindent where $\mathbf{D}$ is a diagonal matrix with element $d_{i,i}$ set to the degree of node $i$, $\lambda_1$ is the largest eigenvalue of $\mathbf{A}$ and $\alpha$ is a flexible parameter constrained to the range $0 < \alpha < 1$.

The association between LHN similarity and the Krylov subspace matrix is revealed by expanding the simplified equation \eqref{eqn:lhn_matrix} as a power series with $\psi$ set to 1:

\begin{equation}\label{eqn:lhn_powseries}
\mathbf{S} = \mathbf{I} + \phi \mathbf{A} + \phi^2 \mathbf{A}^2 + ...
\end{equation}

\noindent As demonstrated by this power series, both the LHN similarity matrix $\mathbf{S}$ and the Krylov subspace matrix $\mathbf{K}$ are defined in terms of the powers of $\mathbf{A}$ whose elements represent the count (or weighted count) of paths of a given length between each pair of nodes. While the LHN similarity for nodes $i$ and $j$ sums across paths of all lengths between the two nodes, the elements of the Krylov subspace trajectory for node $i$ represent the sum across the paths of a specific length between node $i$ and all other nodes. In other words, the Krylov subspace trajectory elements represent the average similarity of a given node to all other nodes where this similarity is computed using only paths of a fixed length.

\section{Network analysis using Krylov subspace trajectories}\label{sec:krylov_applications}

As detailed above, the Krylov subspace trajectories of the adjacency matrix provide important information regarding network structure. In the sections below, we explore three specific network analysis applications of these trajectories:
\begin{enumerate}
\item Node similarity and clustering.
\item Evaluation of network perturbations.
\item Calculation of node importance.
\end{enumerate}

\subsection{Node similarity and clustering using Krylov subspace trajectories}\label{sec:clustering}

One of the key applications of the Krylov subspace trajectories generated by Algorithm \ref{alg:krylov_gen} is the computation of node similarity values, which can be leveraged for clustering/community detection. Importantly, node clustering using trajectory-based similarity values yields very distinct results from standard approaches for network community detection such as Louvain clustering \cite{Blondel_2008} or clustering using LHN similarity values \cite{PhysRevE.73.026120}. To illustrate this phenomenon, the nodes in the tree network visualized in Figure \ref{fig:tree} were clustered using four different methods (results are shown in Figure \ref{fig:tree_clusters}):
\begin{itemize}
\item Louvain clustering as implemented by the \textit{igraph} R package \cite{Csardi:2006aa} function \textit{cluster\_louvain()}.
\item Hierarchical agglomerative clustering (the dendrogram was cut to yield four clusters to match Louvain clustering results) using a distance of one minus the Pearson correlation between Krylov subspace trajectories.
\item Hierarchical agglomerative clustering using a distance of one minus the LHN regular equivalence between nodes (as computed using the \textit{linkprediction} R package function \textit{proxfun(method="lhn\_global")} \cite{linkprediction}). The dendrogram was again cut to yield four clusters.
\item K-means clustering on the eigenvector centrality values with k=4.
\end{itemize}
\noindent For this example, Louvain clustering generates an expected solution that groups nodes according to relative proximity in the graph (Figure \ref{fig:tree_clusters} panel A). Clustering using the Krylov subspace trajectories, on the other hand, generates a very distinct solution that groups nodes based on similar structural position in the network rather than proximity (Figure \ref{fig:tree_clusters} panel B). While the revealed structural-based groups are obvious for the simple tree network, this type of analysis can yield important non-obvious node relationships in the more complex networks used to model real world data. Clustering using LHN similarity values captures elements of both node structural position and node proximity (Figure \ref{fig:tree_clusters} panel C). For this simple network, k-means on the eigenvector centrality values generates the same clustering solution as the trajectory-based approach (Figure \ref{fig:tree_clusters} panel D). The fact that these two approaches yield the same clustering solution follows from the fact that eigenvector centrality values are the last elements of the trajectory vectors and, for this network, trajectories structure is determined by the convergent values. Importantly, the correspondence between trajectory-based node similarity and eigenvector centrality values does not hold in more complex networks (as shown below for a random graph generated according to a preferential attachment model) or when a non-uniform initial vector is used (as explored in Section \ref{sec:perturbation} below).  

\begin{figure}[t]
\begin{center}
\includegraphics[width=0.75\textwidth]{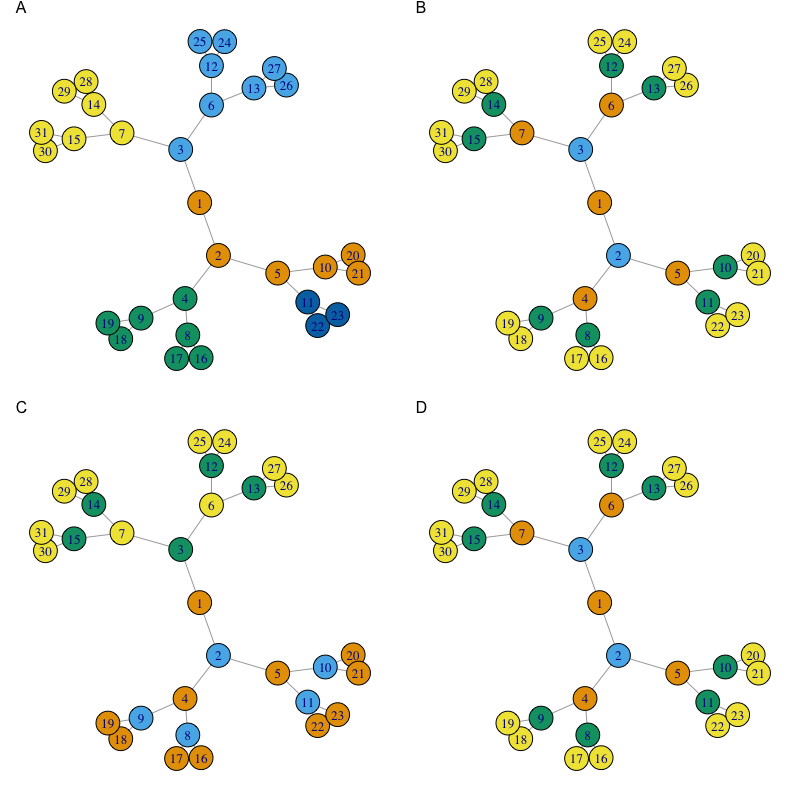}
\end{center}
\caption{Clustering of the example tree network generated by A) Louvain clustering, 
B) hierarchical agglomerative clustering (dendrogram cut at k=4) with correlation distance between the Krylov subspace trajectories, 
C) hierarchical agglomerative clustering using a distance of 1 minus the LHN regular equivalence between nodes, 
and D) k-means clustering on eigenvector centrality values for k=4.}
\label{fig:tree_clusters}
\end{figure}

\clearpage

To capture the results on a more realistic network, a random 50 node network was generated according to the preferential attachment model with $\alpha=1$ using the \textit{igraph} R package function \textit{sample\_pa(n=50,directed=FALSE)}. Clustering of this network was performned using the same four methods applied to the tree network with the number of clusters set to 7 to match the output from Louvain clustering. Similar to the clustering results on the tree network, Louvain clustering of the preferential attachment network captures node proximity, clustering based on Krylov subspace trajectories captures node structural position, and LHN-based clustering reflects a mixture of node proximity and structural position. In contrast to the tree network results, the clusters generated via k-means on the eigenvector centrality values do not match the trajectory-based clusters, i.e., the last element of the trajectory vectors does not determine the rest of the vector elements.

\begin{figure}[t]
\begin{center}
\includegraphics[width=0.75\textwidth]{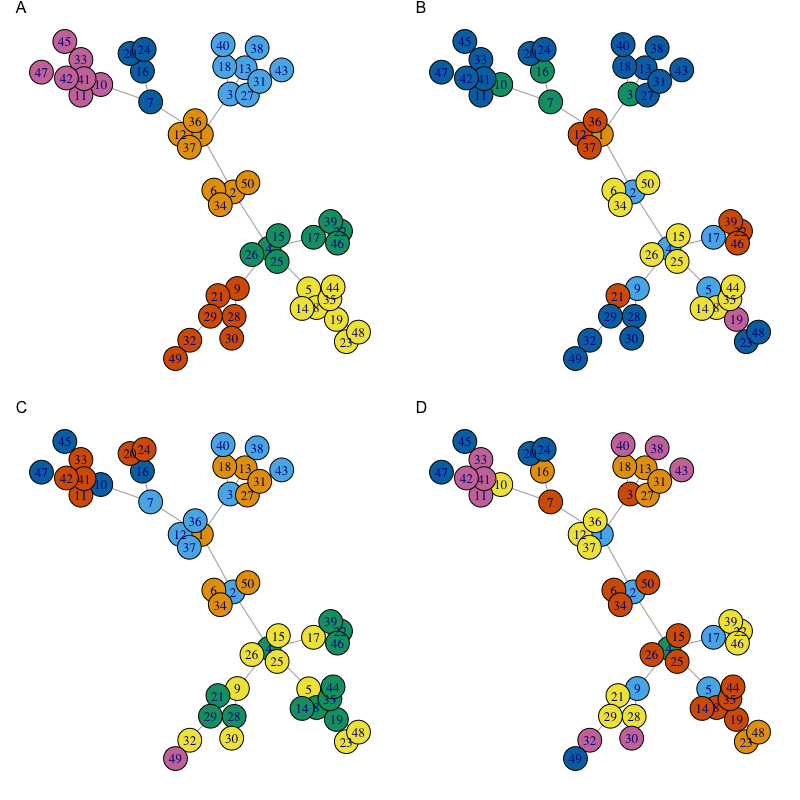}
\end{center}
\caption{Clustering of a random preferential attachment network generated by A) Louvain clustering,
B) hierarchical agglomerative clustering (dendrogram cut at k=7) with correlation distance between the Krylov subspace trajectories,
C) hierarchical agglomerative clustering using a distance of 1 minus the LHN regular equivalence between nodes,
and D) k-means clustering on eigenvector centrality values for k=7.}
\label{fig:tree_clusters}
\end{figure}

\subsection{Perturbation analysis using Krylov subspace trajectories}\label{sec:perturbation}

\begin{figure}[h]
\begin{center}
\includegraphics[width=0.8\textwidth]{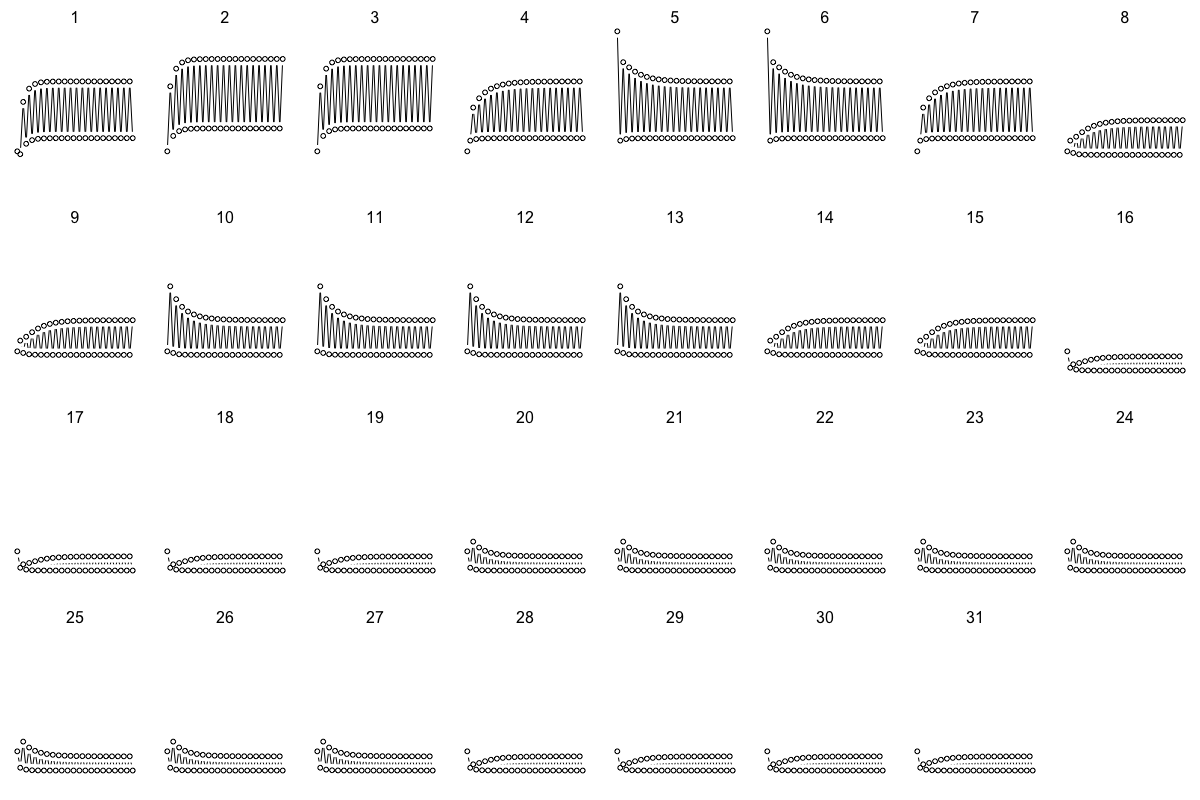}
\end{center}
\caption{Node-specific Krylov trajectories for the tree network visualized in Figure \ref{fig:tree} using a non-uniform $\mathbf{v}_0$ where the entries for nodes 5 and 6 that are five times larger than the entries for other nodes.}
\label{fig:tree_krylov_pert}
\end{figure}

\begin{figure}[t]
\begin{center}
\includegraphics[width=0.8\textwidth]{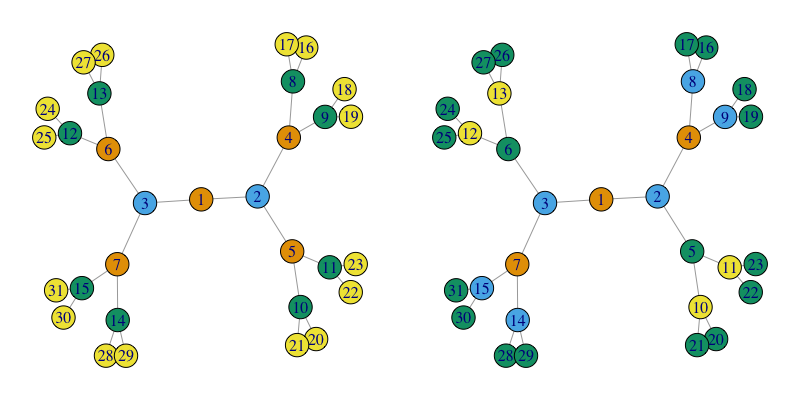}
\end{center}
\caption{Clustering of the example tree network generated by hierarchical agglomerative clustering using correlation distance between the Krylov subspace trajectories and complete linkage. The left panel visualizes the clusters when $\mathbf{v}_0$ is uniform (trajectories are shown in Figure \ref{fig:tree_krylov}) and the right panel visualizes the clusters when $\mathbf{v}_0$ has inflated entries for nodes 5 and 6 and that five times larger than the values for other nodes (trajectories are shown in Figure \ref{fig:tree_krylov_pert}).}
\label{fig:tree_clusters_pert}
\end{figure}

Algorithm \ref{alg:krylov_gen} uses the uniform initial vector $\mathbf{v}_0=\{1/\sqrt{p},...,1/\sqrt{p}\}$ to compute the Krylov subspace trajectories, which is motivated for applications that treat each node equally. For some applications, however, unequal weights can be associated with network nodes and these node weights can be used to create a non-uniform (though still non-random) $\mathbf{v}_0$. One example of such an application would be modeling of a perturbation applied to specific network nodes. To illustrate this type of perturbation analysis, Algorithm \ref{alg:krylov_gen} was applied to the example tree network using a $\mathbf{v}_0$ where the elements for nodes 5 and 6 are five times larger than the elements for other nodes. The trajectories generated in this case are visualized in Figure \ref{fig:tree_krylov_pert}. Clustering based on these perturbed trajectories (shown in the right panel of Figure \ref{fig:tree_clusters_pert}) yields a distinct grouping of nodes that capture both overall structural position and response to the node 5 and 6 perturbations. Importantly, clustering based on the eigenvector centrality values fails to reflect these perturbations, i.e., the solution is generally insensitive to the structure of $\mathbf{v}_0$.

\clearpage

\subsection{Capturing node importance using Krylov subspace trajectories}\label{sec:importance}

\begin{figure}[t]
\begin{center}
\includegraphics[width=0.8\textwidth]{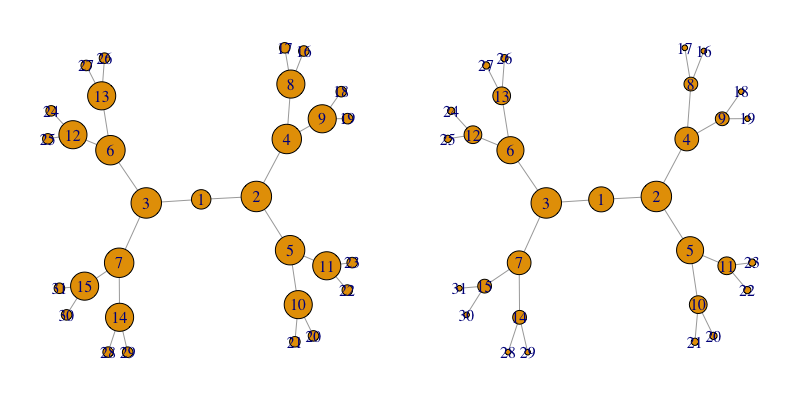}
\end{center}
\caption{Visualization of the $\delta$ statistic for the tree network as computed on the Krylov subspace trajectories for uniform $\mathbf{v}_0$ (left panel; trajectories shown in Figure \ref{fig:tree_krylov}) or non-uniform $\mathbf{v}_0$ (right panel; trajectories shown in Figure \ref{fig:tree_krylov_pert}). To enable visualization for both cases, the $\delta$ values are divided by the maximum $\delta$ for that network.}
\label{fig:tree_delta}
\end{figure}

Krylov subspace trajectories can also be used to capture node importance. Of course, the last value of each trajectory captures eigenvector centrality so we are specifically interested in functions of the entire trajectory that capture aspects of node importance distinct from eigenvector centrality. Although many possible functions exist, we believe that functions which quantify the level of oscillations in the trajectory may have particular utility. One simple way to quantify this is to subtract the absolute difference between the starting and ending values of a given trajectory from the sum of the absolute values of all sequential differences; we refer this trajectory statistic $\delta$. The $\delta$ statistic captures the number and magnitude of oscillations in the trajectory (for monotonic trajectories, it is 0). If the vector $\mathbf{t}$ captures the length $i+1$ trajectory for a given node, then $\delta$ is computed as:
\begin{equation}\label{eqn:extra_delta}
\delta = (\sum_{j=2}^{j=i+1} |t_j-t_{j-1}|) - |t_{i+1}-t_1|
\end{equation}
Figure \ref{fig:tree_delta} visualizes the relative $\delta$ statistics for the example tree network as computed using both uniform and non-uniform $\mathbf{v}_0$.

\section{Conclusions and future directions}
 
In this paper, we detailed a collection of network analysis methods based on Krylov subspace trajectories. These trajectories are defined by the rows of the Krylov subspace matrix computed from a network adjacency matrix via power iteration. While power iteration is commonly applied to adjacency matrices for the computation of eigenvector centrality values, existing approaches use random initial vectors and focus on just a single column of the Krylov subspace matrix (e.g., the vector generated at convergence). Importantly, the use of a non-random initial vector generates non-random trajectories that capture important aspects of network structure and can be leveraged in a range of analysis applications including node clustering, perturbation analysis and computation of node importance. Our future work on this approach will include the analysis of a wider range of simulated networks and evaluation on real networks, e.g., gene regulatory networks.
 
\section*{Acknowledgments}

This work was funded by National Institutes of Health grants R35GM146586, R21CA253408, P20GM130454 and P30CA023108.
We would like to acknowledge the supportive environment at the Geisel School of Medicine at Dartmouth where this research was performed. 


\end{document}